\documentclass[prd,twocolumn,showpacs,preprintnumbers,amsmath,amssymb,nofootinbib]{revtex4}
\pdfoutput=1
\usepackage{amsmath}
\usepackage{graphicx}
\usepackage{dcolumn}
\usepackage{xspace}
\usepackage{color}
\usepackage{url}
\usepackage[utf8]{inputenc}
\usepackage{epstopdf}

\newcommand{\bea}{\begin{eqnarray}}
\newcommand{\eea}{\end{eqnarray}}
\newcommand{\nn}{\nonumber\\}

\newcommand{\eq}[1]{Eq.~\eqref{#1}}

\begin{document}
\preprint{PSI-PR-17-04, YACHAY-PUB-17-03-PN, ZU-TH 05/17}
\title{Simultaneous Explanation of $R(D^{(*)})$ and $b\to s\mu^+\mu^-$: \\The Last Scalar Leptoquarks Standing}

\author{Andreas Crivellin}
\email{andreas.crivellin@cern.ch}
\affiliation{Paul Scherrer Institut, CH--5232 Villigen PSI, Switzerland}

\author{Dario M\"uller}
\email{dario.mueller@psi.ch}
\affiliation{Paul Scherrer Institut, CH--5232 Villigen PSI, Switzerland}
\affiliation{Physik-Institut, Universit\"at Z\"urich,
Winterthurerstrasse 190, CH-8057 Z\"urich, Switzerland}

\author{Toshihiko Ota}
\email{tota@yachaytech.edu.ec}
\affiliation{%
Department of Physics, Yachay Tech,
Hacienda San Jos\'{e} s/n y Proyecto Yachay, 
100119 San Miguel de Urcuqu\'{i}, 
Ecuador}

\begin{abstract}
Over the past years, experiments accumulated intriguing hints for new
 physics (NP) in flavor observables, namely in the anomalous magnetic
 moment of the muon ($a_\mu$), in $R(D^{(*)})={\rm Br}(B\to
 D^{(*)}\tau\nu)/{\rm Br}(B\to D^{(*)}\ell\nu)$ and in $b\to
 s\mu^+\mu^-$ transitions, which are all at the $3-4\,\sigma$ level. In
 this article we point out that one can explain the $R(D^{(*)})$ anomaly
 using two scalar leptoquarks (LQs) with the same mass and coupling to
 fermions related via a discrete symmetry: an $SU(2)_L$ singlet and an
 $SU(2)_L$ triplet, both with hypercharge $Y=-2/3$. In this way,
 potentially dangerous contributions to $b\to s\nu\nu$ are avoided and
 non-CKM suppressed effects in $R(D^{(*)})$ can be generated. This
 allows for smaller overall couplings to fermions weakening the direct
 LHC bounds. In our model, $R(D^{(*)})$ is directly correlated to $b\to
 s\tau^+\tau^-$ transitions where an enhancement by orders of magnitude
 compared to the standard model (SM) is predicted, such that these decay modes are in the reach of LHCb and BELLE II. Furthermore, one can also naturally explain the $b\to s\mu^+\mu^-$ anomalies (including $R(K)$) by a $C_9=-C_{10}$ like contribution without spoiling $\mu-e$ universality in charged current decays. In this case sizable effects in $b\to s\tau\mu$ transitions are predicted which are again well within the experimental reach. One can even address the longstanding anomaly in $a_\mu$, generating a sizable decay rate for $\tau\to\mu\gamma$. However, we find that out of the three anomalies $R(D^{(*)})$, $b\to s\mu^+\mu^-$ and $a_{\mu}$ only two (but any two) can be explained simultaneously. We point out that a very similar phenomenology can be achieved using a vector leptoquark $SU(2)_L$ singlet with hypercharge $2/3$. In this case, no tuning between couplings is necessary, but the model is non-renormalizable.
\end{abstract}

\maketitle

\section{Introduction}
\label{intro}

So far, the LHC did not directly observe any particles beyond the ones present in the SM of particle physics. However, we have intriguing hints for lepton flavor universality violating NP\footnote{See for example Ref.~\cite{Crivellin:2016ekz} for a recent overview.}. Most prominently, there exist deviations from the SM predictions in $b\to s\mu^+\mu^-$ at the $4-5\,\sigma$ level \cite{Altmannshofer:2015sma,Descotes-Genon:2015uva,Hurth:2016fbr} and the combination of the ratios $R(D)$ and $R(D^*)$ differs by $3.9\,\sigma$ from its SM prediction~\cite{Amhis:2016xyh}. Furthermore, the longstanding anomaly in $a_\mu$ ($3.1\,\sigma$~\cite{Nyffeler:2016gnb}) also points towards NP.

$R(D)$ and $R(D^*)$ directly measure lepton flavor universality violation (LFUV), and in the fit to the $b\to s\mu^+\mu^-$ data also the LHCb measurement of $R(K)$~\cite{Aaij:2014ora}, which deviates by $2.6\,\sigma$ from the SM, points at LFUV. Therefore, it is well motivated to search for a simultaneous explanation of these two anomalies~\cite{Bhattacharya:2014wla,Calibbi:2015kma,Fajfer:2015ycq,Greljo:2015mma,Barbieri:2015yvd,Bauer:2015knc,Boucenna:2016qad,Das:2016vkr,Becirevic:2016yqi,Sahoo:2016pet,Bhattacharya:2016mcc,Barbieri:2016las,Chen:2017hir}. Furthermore, since $a_e$ agrees with the SM prediction, also $a_\mu$ can be considered as a LFUV quantity and one can address it together with $R(D^{(*)})$ and/or $b\to s\mu^+\mu^-$~\cite{Crivellin:2015hha,Belanger:2015nma,Bauer:2015knc,Altmannshofer:2016oaq}.

\begin{figure*}[t]
\begin{center}
	\begin{tabular}{cp{7mm}c}
		\includegraphics[width=0.9\textwidth]{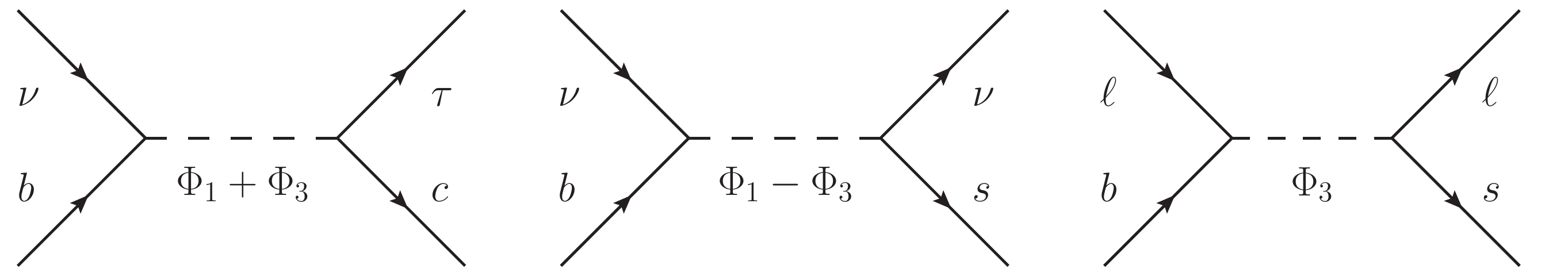}
	\end{tabular}
\end{center}
\caption{Feynman diagrams contributing to $b\to c\tau\nu$, $b\to s\nu\nu$ and $b\to s\ell\ell$ processes. Both LQs contribute to $b\to c\tau\nu$ and $b\to s\nu\nu$ but only $\Phi_3$ to $b\to s\ell\ell$. Note that with our assumption on the couplings to fermions, the LQs interfere constructively (destructively) in $b\to c\tau\nu$ ($b\to s\nu\nu$).} 
\label{FeynmanDiagrams}
\end{figure*}

Concerning $b\to s\mu^+\mu^-$, a solution is not particularly challenging, as one competes with a process which is in the SM loop and CKM suppressed, a rather small NP contribution, involving moderate couplings and not too light masses, is sufficient (like for example in $Z^\prime$ models~\cite{Descotes-Genon:2013wba,Gauld:2013qba,Buras:2013qja,Buras:2013dea,Altmannshofer:2014cfa,Crivellin:2015mga,Crivellin:2015lwa,Niehoff:2015bfa,Sierra:2015fma,Crivellin:2015era,Celis:2015ara,Crivellin:2016ejn,GarciaGarcia:2016nvr}, models with loop effects of heavy scalars and fermions~\cite{Gripaios:2015gra,Arnan:2016cpy} and also by leptoquark models~\cite{Gripaios:2014tna,Fajfer:2015ycq,Becirevic:2015asa,Varzielas:2015iva,Alonso:2015sja,Calibbi:2015kma,Barbieri:2015yvd}\footnote{In Ref.~\cite{Jager:2017gal} it was pointed out that NP effects in charm operators could account for the anomalies. However, this would lead to $q^2$ dependent effects and it could not explain signs for LFUV.}). Also for the $a_\mu$ anomaly many possible solutions exist. Here, we would just like to stress that LQs provide a natural solution since they can give the desired large effect because of an $m_t/m_\mu$ enhancement~\cite{Djouadi:1989md,Chakraverty:2001yg,Cheung:2001ip,Bauer:2015knc,ColuccioLeskow:2016dox}. 

However, an explanation of $R(D)$ and $R(D^*)$ is getting more and more delicate. Since these processes are mediated in the SM already at tree-level, a rather large NP contribution is required to account for the $O(20\%)$ deviation. Therefore, new particles added to the SM for explaining $R(D)$ and $R(D^*)$ cannot be very heavy and must have sizable couplings. In the past, mainly three kinds of models with the following new particles have been proposed:
\begin{enumerate}
  \vspace{-2mm}
	\item Charged Higgses~\cite{Crivellin:2012ye,Tanaka:2012nw,Celis:2012dk,Crivellin:2013wna,Crivellin:2015hha,
	Chen:2017eby}
	\vspace{-2mm}
	\item $W^\prime$ gauge bosons~\cite{Greljo:2015mma,Boucenna:2016wpr,Boucenna:2016qad,Megias:2017ove}
	\vspace{-2mm}
	\item Leptoquarks~\cite{Fajfer:2012jt,Deshpande:2012rr,Sakaki:2013bfa,Alonso:2015sja,Calibbi:2015kma,Bauer:2015knc,Fajfer:2015ycq,Barbieri:2015yvd,Deshpand:2016cpw,Li:2016vvp,Becirevic:2016yqi,Dumont:2016xpj,Das:2016vkr,Barbieri:2016las,Chen:2017hir}
	\vspace{-2mm}
\end{enumerate}
Models with charged Higgses lead to (too) large effects in the total $B_c$ lifetime~\cite{Alonso:2016oyd} and, depending on the coupling structure, can also be disfavored by the $q^2$ 
distribution~\cite{Freytsis:2015qca,Celis:2016azn,Ivanov:2017mrj}. Interestingly, if the couplings of the charged Higgs are chosen in such a way that they are compatible with the measured $q^2$ distribution, these models are ruled out by direct searches~\cite{Faroughy:2016osc}.

Models with $W^\prime$ gauge bosons are also delicate because they necessarily involve $Z^\prime$ bosons due to $SU(2)_L$ gauge invariance. If the $Z^\prime$ width is not unnaturally large, these models are again ruled out by direct searches~\cite{Greljo:2015mma,Faroughy:2016osc}.

In models with leptoquarks generating left-handed vector operators the coupling structure should be aligned to the bottom quark in order to avoid $b\to s\nu\nu$ bounds. However, in this case the effect in $R(D)$ and $R(D^*)$ is proportional to the small CKM element $V_{cb}$ and large third generation couplings are required to account for the anomalies. These large third generation couplings lead again to stringent bounds from direct LHC searches~\cite{Faroughy:2016osc} and electroweak precision observables~\cite{Feruglio:2016gvd}. In principle, these constraints can be avoided with right-handed couplings~\cite{Li:2016vvp} (including possibly right-handed neutrinos~\cite{Becirevic:2016yqi}). However, in such solutions no interference with the SM appears and very large couplings, close to non-perturbativity, are required. 
 
\begin{figure*}[t]
\begin{center}
\begin{tabular}{cp{7mm}c}
\includegraphics[width=0.4125\textwidth]{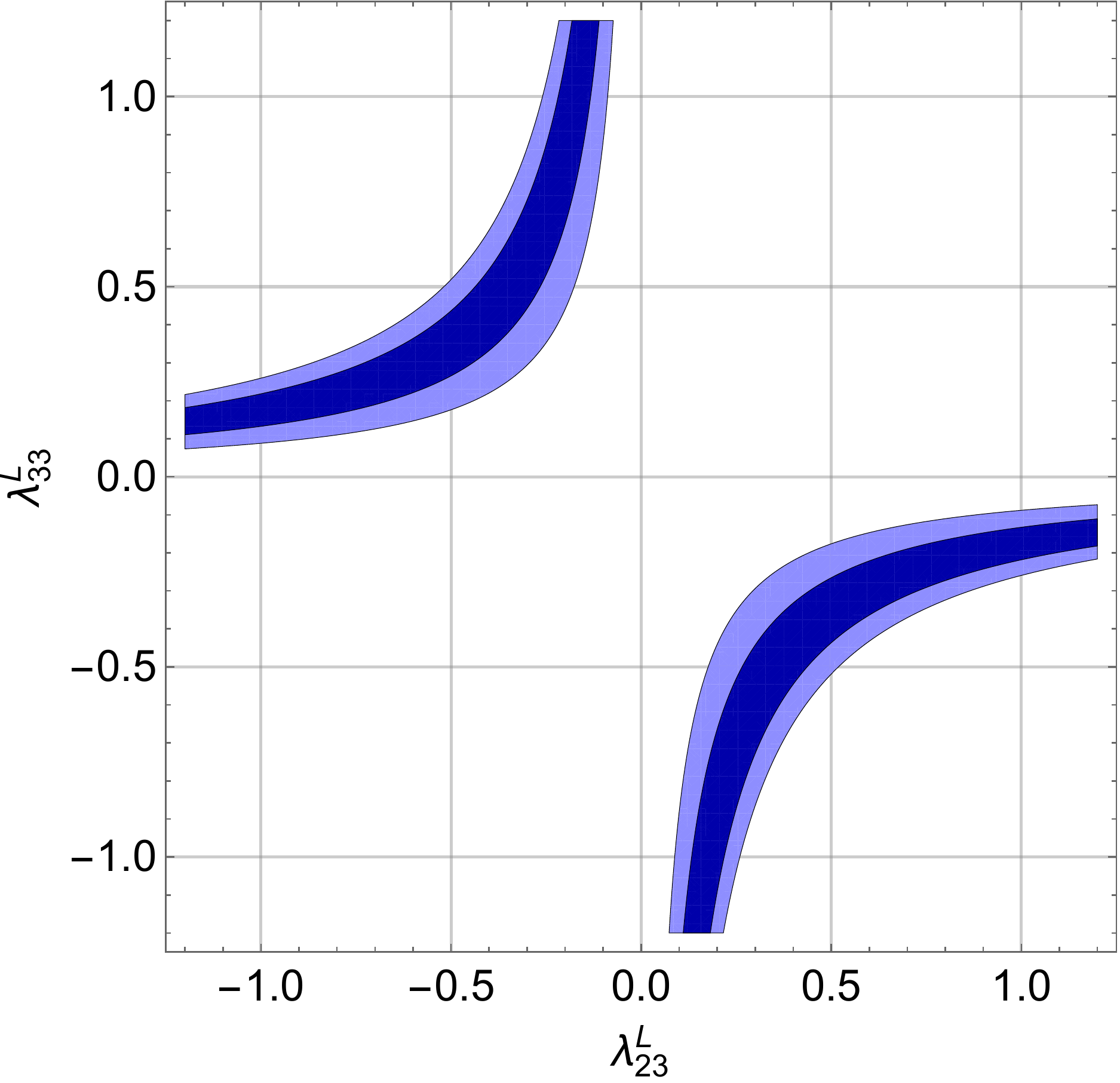}
\includegraphics[width=0.58\textwidth]{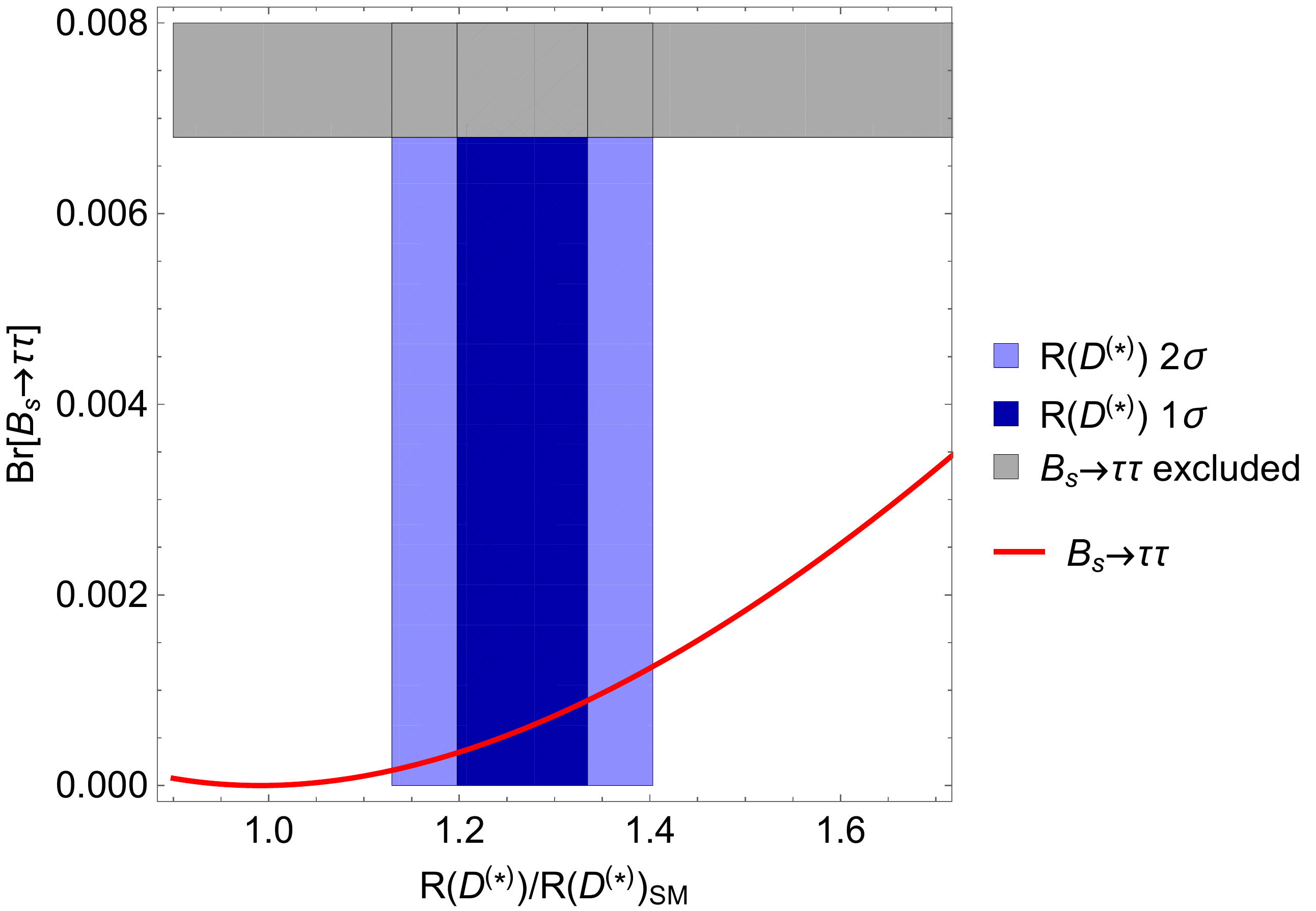}
\end{tabular}
\end{center}
\caption{Left: Allowed regions by $R(D)$ and $R(D^*)$ in the $\lambda^L_{23}-\lambda^L_{33}$ plane for $M=1\,$TeV using the weighted average for $R(D^{(*)})_{\rm EXP}/R(D^{(*)})_{\rm SM}$. Note that already small couplings are sufficient to account for $R(D)$ and $R(D^*)$. Therefore, the bounds from LHC searches are weakened and the leptoquarks can also be easily heavier than $1\,$TeV and still explain the anomalies with couplings in the perturbative regime. Right: Prediction for $B_s\to\tau\tau$ (red) as a function of $R(D^{(*)})/R(D^{(*)})_{\rm SM}$. Here we neglected small CKM suppressed contributions.}         
\label{Bstautau}
\end{figure*}

As stated above, LHC bounds from $\tau\tau$ searches can be avoided in
case of non-CKM suppressed leptoquark contributions to $R(D)$ and
$R(D^*)$. However, for single scalar leptoquark representations, this
leads to unacceptably large effects in $b\to s$
transitions~\cite{Li:2016vvp}. Therefore, we propose a novel solution to
the $R(D^{(*)})$ problem in this article: we introduce two scalar
leptoquarks with the same mass $M$ and the same coupling strength to
quarks and leptons; an $SU(2)_L$ singlet ($\Phi_1$) and an $SU(2)_L$ triplet
($\Phi_3$) both with hypercharge $Y=-2/3$. Here, the crucial observation
is that $\Phi_1$ and $\Phi_3$ contribute with opposite relative sign to
$R(D^{(*)})$ than to $b\to s\nu\nu$ processes such that the
effect in $R(D^{(*)})$ is doubled while the contributions in $B\to
K^{(*)}\nu\nu$ cancel at tree-level (see
Fig.~\ref{FeynmanDiagrams}). Therefore, the couplings to the second
quark generation can be larger, non-CKM suppressed effects $R(D^{(*)})$
are possible and the required overall coupling strength is much smaller
such that the direct LHC bounds from $\tau\tau$ searches are significantly weakened and the remaining bounds from pair production of third generation LQs are still below the TeV scale~\cite{Khachatryan:2016ycy,Aad:2016kww}. Furthermore, this solution results in a simple rescaling of the SM contributions, predicts naturally $R(D)/R(D)_{\rm SM}=R(D^*)/R(D^*)_{\rm SM}$ and leaves the $q^2$ distribution unchanged. Adding couplings to muons, we can also address the $b\to s\mu\mu$ anomalies with a $C_9=-C_{10}$ like contribution. Finally, adding a (small) right-handed coupling of $\Phi_1$ one can in principle explain $a_\mu$. 

This article is structured as follows: in the next section we will present the contributions of our model to all relevant observables. Afterwards, we perform a phenomenological analysis in Sec.~\ref{analysis} before we conclude.

\section{Model and observables}

The scalar leptoquark singlet $\Phi_1$ and the triplet $\Phi_3$ couple to fermions in the following way:
\begin{align}
L = \lambda _{fi}^{1L}\overline {Q_f^c} i{\tau _2}{L_i}\Phi _1^\dag  + \lambda _{fi}^{3L}\overline {Q_f^c} i{\tau _2}{\left( {\tau  \cdot \Phi _3^{}} \right)^\dag }{L_i} + {\rm{h}}{\rm{.c}}.\,.\label{LLQ}
\end{align}
As motivated in the introduction, we assume that both leptoquarks have the same mass $M$. In addition, to cancel their effect in $b\to s\nu\nu$ processes, we impose the discrete symmetry
\begin{equation}
	\lambda _{jk}^{L}\equiv\lambda _{jk}^{1L}\,,\qquad   \lambda_{jk}^{3L}=e^{i\pi j}\lambda _{jk}^{L}\,,
\end{equation}
on the couplings to fermions. Note that for $\Phi_1$ there is in principle an additional coupling $\lambda _{fi}^R\overline {u_f^c} {\ell _i}\Phi _1^\dag$ allowed. We will assume that this coupling is zero and neglect its effect till the discussion of $a_\mu$ where small values of $\lambda_{fi}^R$ can be phenomenologically important due to $m_t/m_\mu$ enhanced effects. For our analysis we assume that the couplings $\lambda_{fi}^L$ are given in the down-quark basis. I.e. after EW symmetry breaking the couplings to left-handed up-quarks involve CKM elements:
\begin{equation}
	\lambda _{{d_f}i}^L \equiv \lambda _{fi}^L,\;\;\;\lambda _{{u_f}i}^L = V_{fj}^*\lambda _{ji}^L\,.
\end{equation}

We will now discuss the various relevant processes to which our model contributes. As already noted above, our model is constructed in such a way that we do not get tree-level contributions to $b\to s\nu\nu$ transitions, which we therefore omit in the following. We also neglect couplings to the first generation of quarks and leptons. For quarks, this is only possible in the interaction basis since CKM rotations induce either couplings to up or down quarks. However, charged current decays of Kaons or $D$ mesons involve large CKM angles in the SM, making the relative effects of LQs small.

\subsection{$R(D)$ and $R(D^*)$}

We define the effective Hamiltonian for $b\to c\ell\nu$ transitions as
\begin{equation}
{H_{{\rm{eff}}}^{\ell_f\nu_i}} = \frac{{4{G_F}}}{{\sqrt 2 }}{V_{cb}}C_L^{fi}\left[ {\bar c{\gamma ^\mu }{P_L}b} \right]\left[ {{{\bar \ell }_f}{\gamma _\mu }{P_L}{\nu _i}} \right]\,,
\end{equation}
where in the SM $C_L^{fi}=\delta_{fi}$ and the contribution of our model is given by
\begin{equation}
	C_L^{fi} = \frac{{ \sqrt 2 }}{{8{G_F}{M^2}}}\frac{{V_{cj}}}{{V_{cb}^{}}}\lambda _{3i}^L\lambda _{jf}^{L*}\left(1+(-1)^{j}\right)\,.
\end{equation}
With these conventions we have
\begin{equation}
{{R\left( {{D^{\left( * \right)}}} \right)} \mathord{\left/
 {\vphantom {{R\left( {{D^{\left( * \right)}}} \right)} {R{{\left( {{D^{\left( * \right)}}} \right)}_{{\rm{SM}}}}}}} \right.
 \kern-\nulldelimiterspace} {R{{\left( {{D^{\left( * \right)}}} \right)}_{{\rm{SM}}}}}} \equiv X_{D^{(*)}}= \sum\limits_{i = 1}^3 {{{\left( {{\delta _{3i}} + C_L^{3i}} \right)}^2}}\,,
\end{equation}
assuming vanishing contributions to the muon and electron channels. This has to be compared to the experimental values of
\begin{eqnarray}
R{\left( {{D^*}} \right)_{{\rm{EXP}}}} &=& 0.316 \pm 0.016 \pm 0.010\,,\\
R{\left( D \right)_{{\rm{EXP}}}} &=& 0.397 \pm 0.040 \pm 0.028\,,
\end{eqnarray}
and the corresponding SM predictions~\cite{Fajfer:2012vx,Na:2015kha}
\begin{eqnarray}
R{\left( {{D^*}} \right)_{{\rm{SM}}}} &=& 0.252 \pm 0.003\,,\\
R{\left( D \right)_{{\rm{SM}}}} &=& 0.300 \pm 0.008\,.
\end{eqnarray}
%


\subsection{$b\to s\ell^+\ell^-$ transitions}

Using the effective Hamiltonian
\begin{align}
H_{\rm eff}^{\ell_f\ell_i}=- \dfrac{ 4 G_F }{\sqrt 2}V_{tb}V_{ts}^{*} \sum\limits_{a = 9,10} C_a^{fi} O_a^{fi}\,
,\nn {O_{9(10)}^{fi}} =\dfrac{\alpha }{4\pi}[\bar s{\gamma ^\mu } P_L b]\,[\bar\ell_f{\gamma _\mu }(\gamma^5)\ell_i] \,,
\label{eq:effHam}
\end{align}
we have
\begin{equation}
C_9^{fi} =  - C_{10}^{fi} = \frac{{ -\sqrt 2} }{{2{G_F}{V_{tb}}V_{ts}^*}}\frac{\pi }{\alpha }\frac{1}{{{M^2}}}\lambda _{3i}^{L}\lambda _{2f}^{L*}\,.
\end{equation}
The allowed range at the $2\,\sigma$ level~\cite{Altmannshofer:2015sma} (see also Ref.~\cite{Descotes-Genon:2015uva,Hurth:2016fbr}) is given by
\begin{equation}
	-0.18(-0.35) \geq  C_{9}^{22}=-C_{10}^{22} \geq (-0.71) -0.91\,,
\end{equation}
at the ($1\,\sigma$) $2\,\sigma$ level.

We will also need the process $B_s\to\tau^+\tau^-$. The current experimental limit is~\cite{Aaij:2017xqt}
\begin{equation}
{\rm{Br}}{\left( {B_s \to {\tau ^ + }{\tau ^ - }} \right)_{{\rm{EXP}}}} \le 6.8 \times {10^{ - 3}}\,.
\end{equation}
The SM prediction is given by~\cite{Bobeth:2013uxa,Bobeth:2014tza}
\begin{equation}
	{\rm{Br}}{\left( {B_s \to {\tau ^ + }{\tau ^ - }} \right)_{{\rm{SM}}}} = \left( {7.73 \pm 0.49} \right) \times {10^{ - 7}}\,,
\end{equation}
and in our model we have
\begin{equation}
{\rm{Br}}\left( {{B_s} \to {\tau ^ + }{\tau ^ - }} \right) =
 {\rm{Br}}{\left( {{B_s} \to {\tau ^ + }{\tau ^ - }} \right)_{\rm
 SM}}{\left( {1 + \frac{{C_{10}^{33}}}{{C_{10}^{\rm SM}}}} \right)^2}\,,
\end{equation}
with $C_{10}^{\rm SM} \approx  - 4.3$~\cite{Bobeth:1999mk,Huber:2005ig}. 
For the analysis of $B\to K^{(*)}\tau\mu$ we will use the results of Ref.~\cite{Crivellin:2015era}.

\subsection{$B^{0}-\bar{B}^{0}$ mixing}

Here we find
\begin{equation}
\begin{array}{l}
{H_{{\rm{eff}}}} = {C_1}\bar s{\gamma ^\mu }{P_L}b\bar s{\gamma _\mu }{P_L}b\\
{C_1} = \frac{{ - 1}}{{128{\pi ^2}}}{\left( {\lambda _{23}^{L*}\lambda _{33}^L} \right)^2}{D_2}\left( {m_\tau ^2,m_\tau ^2,{M^2},{M^2}} \right)
\end{array}
\end{equation}
which corresponds to an effect of the order of 1\%. This can be easily understood as follows: Since we need an $\mathcal{O}(10\%)$ effect in $R(D^{(*)})$ at the amplitude level, this effects gets squared for $B^{0}-\bar{B}^{0}$ and there are no enhancement factors, the final effect is around 1\% and below the current sensitivity of approximately 10\%.

\subsection{$a_\mu$ and $\tau\to\mu\gamma$}

In order to aim at an explanation of $a_\mu$ one needs a chirality enhanced effect. Therefore, let us add to \eq{LLQ} the following term
\begin{equation}
\lambda_{fi}^R\overline{u^c_f}\ell_i\Phi_1^{\dagger}+h.c.\,. \label{lambdaR}
\end{equation}
In this case the numerically relevant $m_t$ enhanced contribution to $a_\mu$ is given by
\begin{equation}
{\delta a_\mu } = \frac{{{m_\mu }}}{{4{\pi ^2}}}{\mathop{\rm Re}\nolimits} \left[ {C_R^{22}} \right]\,,
\end{equation}
with
\begin{eqnarray}
C_L^{fi} =  - \frac{{{N_c}}}{{12{M^2}}}{m_t}\lambda _{3f}^R\lambda _{3i}^{L*}\left( {7 + 4\log \left( {\frac{{m_t^2}}{{{M^2}}}} \right)} \right)\,,
\end{eqnarray}
and $C_{R}^{23}$ is obtained from $C_{L}^{23}$ by $L \leftrightarrow R$. We will assume that $\lambda _{32}^R$ is small compared to $\lambda _{32}^L$.

The world average of the measurement of $a_\mu \equiv (g-2)_\mu/2$ is completely dominated by the Brookhaven experiment E821~\cite{Bennett:2006fi} and is given by~\cite{Agashe:2014kda}
$a_\mu^\mathrm{exp} = (116\,592\,091\pm54\pm33) \times 10^{-11}$
where the first error is statistical and the second one is systematic. The current SM prediction is~\cite{Czarnecki1995,Czarnecki1996,Gnendiger2013,Davier2011,Hagiwara2011,Kurz2014,Jegerlehner2009,Aoyama2012,Colangelo2014,Nyffeler:2016gnb}
$	a_\mu^\mathrm{SM} = (116\,591\,811\pm62) \times 10^{-11}$ where almost the whole uncertainty is due to hadronic effects. This amounts to a discrepancy between the SM and the experimental value of 
\begin{equation}
\delta a_\mu=	a_\mu^\mathrm{exp}-a_\mu^\mathrm{SM} = (278\pm 88)\times 10^{-11}\,,\label{muonAMMexp}
\end{equation}
i.e.~a $3.1\sigma$ deviation.

For $\tau\to\mu\gamma$ the branching ratio reads
\begin{equation}
{\rm{Br}}\left( {{\tau} \to {\mu}\gamma } \right) = \frac{{{\alpha}{m_{{\tau}}^3}}}{{256{\pi^4}}}{\tau _{{\tau}}}\left( {{{\left| {C_L^{23}} \right|}^2} + {{\left| {C_R^{23}} \right|}^2}} \right)\,.
\end{equation}
The current experimental bound is given by~\cite{Aubert:2009ag}
\begin{equation}
	{\rm Br}(\tau \to \mu \gamma) < 4.4 \times 10^{-8}\,.
\end{equation}

\begin{figure*}[t]
\begin{center}
\begin{tabular}{cp{7mm}c}
\includegraphics[width=0.55\textwidth]{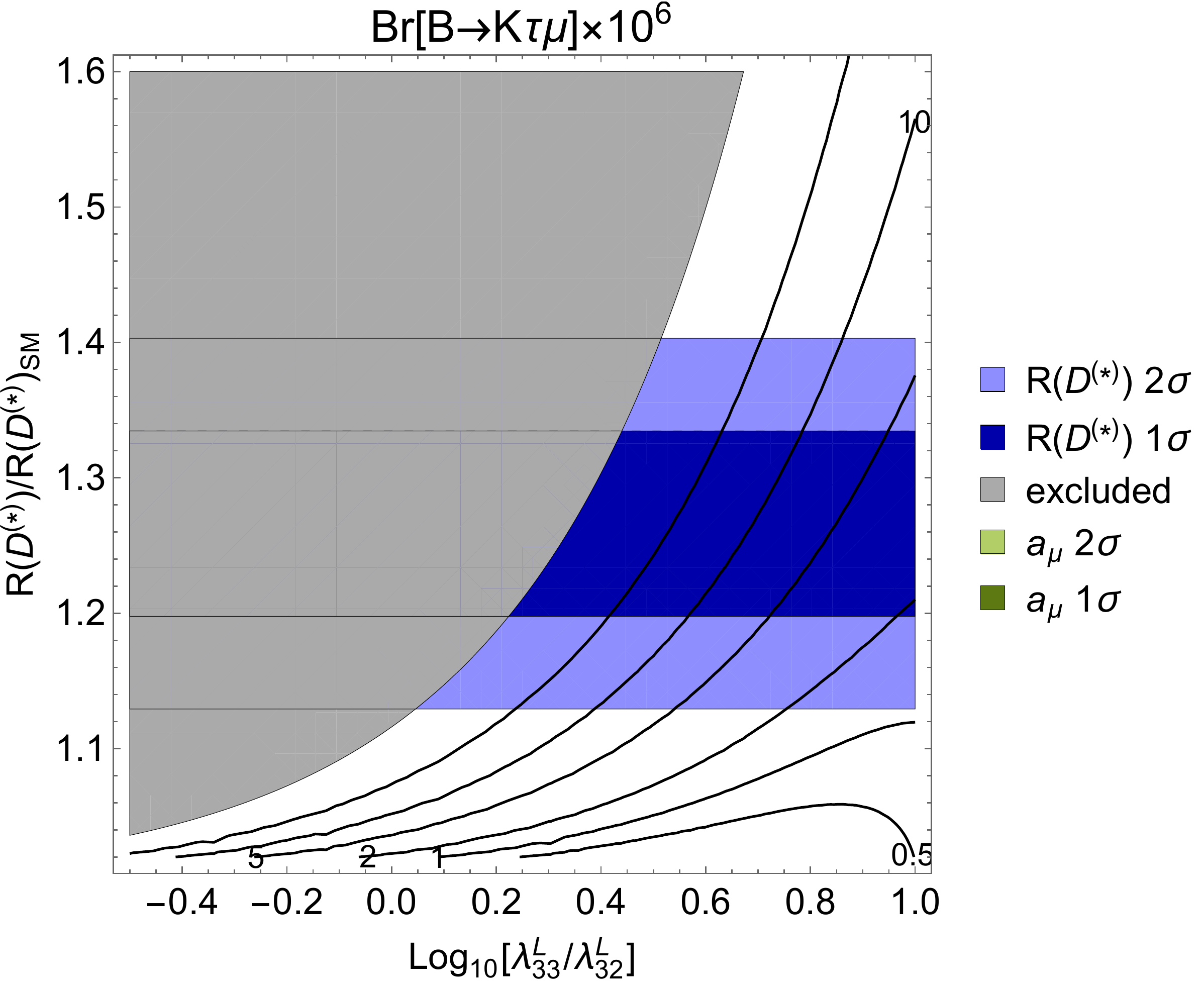}
\includegraphics[width=0.4375\textwidth]{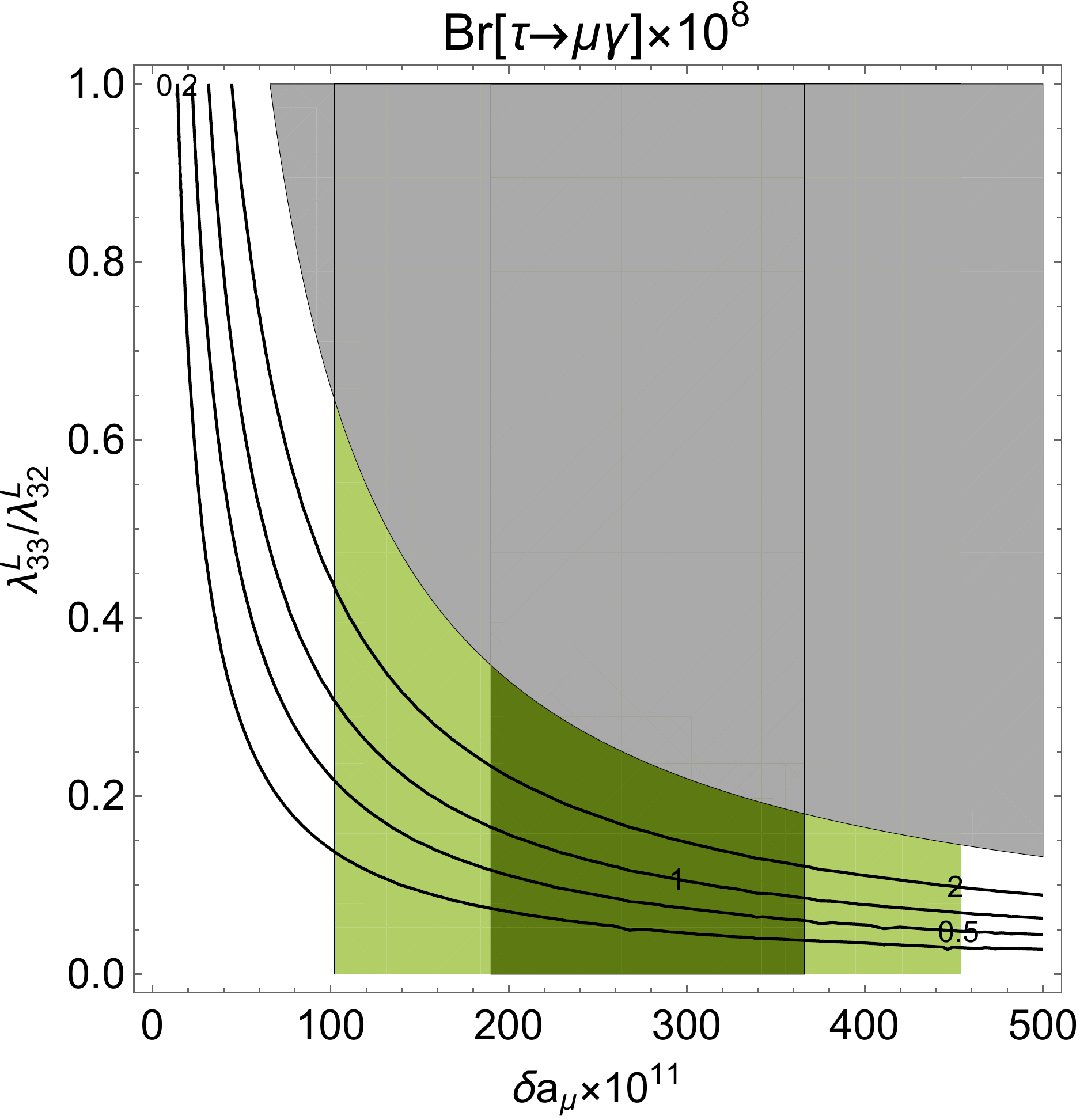}
\end{tabular}
\end{center}
\caption{Left: Contours and excluded region for $B\to K\tau\mu=(B\to K\tau^+\mu^-+B\to K\tau^-\mu^+)/2$ for $C_9^{22}=-0.5$, i.e. assuming that $C_{9}^{22}$ takes the central value obtained from the $b\to s\mu^+\mu^-$ fit. The colored regions are allowed by the various processes. For $R(D)$ and $R(D^*)$ we used again the weighted average for $R(D^{(*)})_{\rm EXP}/R(D^{(*)})_{\rm SM}$. Right: The contour lines show ${\rm Br}[\tau\to\mu\gamma]\times 10^{8}$. The gray region is excluded by the current upper limit and (light) red region is allowed by $a_\mu$ at the ($2\,\sigma$) $1\,\sigma$ level. Note that both $\delta a_{\mu}$ and $\tau\to\mu\gamma$ are only a function of $\lambda_{33}^L/\lambda_{32}^L$ and therefore independent of $b\to s\mu^+\mu^-$ transitions.}   
\label{BKtaumuAMM}
\end{figure*}

\section{Phenomenological analysis}
\label{analysis}

\subsection{$R(D)$, $R(D^*)$ and $b\to s\tau^+\tau^-$}

Let us first consider the size of the couplings needed to explain $R(D)$ and $R(D^*)$. Here and in the following, we will assume them to be real. As we can see in the left plot of Fig.~\ref{Bstautau}, we only need small couplings (of the order of $0.1$ for 1 TeV leptoquarks) in order to explain $R(D)$, $R(D^*)$. This is possible because we avoid contributions to $b\to s\nu\nu$ and hence our effect in $b\to c\tau\nu$ does not need to be CKM suppressed. Therefore, the bounds from Ref.~\cite{Faroughy:2016osc} do not apply to our model and we are not in conflict with LHC bounds, especially because the LQs can be much heavier than 1 TeV while still possessing perturbative couplings and explaining $R(D^{(*)})$.

Next, note that neglecting small CKM factors, the contributions to $b\to c\tau\nu$ and $b\to s\tau\tau$ depend on the same product of couplings $\lambda^L_{23}\lambda^{L*}_{33}$ (modulus small CKM ratios). Therefore, we can express $B_s\to\tau^+\tau^-$ in terms of the effect in $R(D^{(*)})$:
\begin{equation}
\frac{{{\rm{Br}}\left( {{B_s} \to \tau \tau }
		\right)}}{{{\rm{Br}}{{\left( {{B_s} \to \tau \tau }   \right)}_{\rm SM}}}} =
		\left(	 {1 +		2\frac{\pi }{\alpha	}\frac{{V_{cb}^{}}}{{V_{ts}^*}}\frac{{\sqrt
		{{X_{{D^{\left( * \right)}}}}}  - 1}}{{C_{10}^{\rm
		SM}}}}
		\right)^{2}\,.
\end{equation}
The resulting numerical prediction for $B_s\to\tau^+\tau^-$ is  shown in Fig.~\ref{Bstautau}. We can see that the branching ratio can be enhanced by up to three orders of magnitude compared to the SM prediction. Therefore, even though it is experimentally challenging to search for, our model can be tested with $B_s\to\tau^+\tau^-$ measurements at LHCb. Also an enhancement of $B\to K^{(*)}\tau^+\tau^-$ in the same ballpark is predicted which could be tested at BELLE II.

\subsection{$b\to s\mu^+\mu^-$ and $b\to s\tau^\pm\mu^\mp$}

Let us now consider the effect of including $b\to s\mu^+\mu^-$ transitions in our analysis. In this case effects in $B\to D^{(*)}\mu\nu/B\to D^{(*)}e\nu$ are predicted if still addressing $R(D)$ and $R(D^*)$ simultaneously. We checked that the effect is at the per-mill level which is compatible with BELLE and BABAR measurements\footnote{This is contrary to Ref.~\cite{Bauer:2015knc} which cannot explain $R(D^{(*)})$ and $b\to s\mu^+\mu^-$ data simultaneously without violating the bounds from $B\to D^{(*)}\mu\nu/B\to D^{(*)}e\nu$ as pointed out in Ref.~\cite{Becirevic:2016oho}. However, this tension can be relieved with leptoquarks masses larger than $5$ TeV \cite{Cai:2017wry}.}.
However, interesting correlations with $b\to s\tau\mu$ processes appear. Here we find
\begin{eqnarray}
C_9^{32} &=& -2\frac{\pi }{\alpha }\frac{{V_{cb}^{}}}{{V_{ts}^*}}\frac{\lambda^{L}_{32}}{\lambda^{L}_{33}}\left( {\sqrt {{X_{{D^{\left( * \right)}}}}}  - 1} \right)\,,\\
C_9^{23} &=&  \frac{{\lambda _{33}^L}}{{\lambda _{32}^L}}C_9^{22}\,,
\end{eqnarray}
which depends only on the ratio ${{\lambda _{33}^L}}/{{\lambda _{32}^L}}$ as a free parameter. Note that the dependence on $C_9^{22}$ is much weaker than on $X_{{D^{\left( * \right)}}}$. The resulting bounds and predictions are shown in the left plot of Fig.~\ref{BKtaumuAMM}. We take the experimental limit \cite{Lees:2012zz}
\begin{align}
\rm{Br}\left[B\to K\tau\mu\right]<4.8\times 10^{-5}\,.
\end{align}
Note that $R(D^{(*)})$ can only be fully explained for ${{\lambda _{33}^L}}/{{\lambda _{32}^L}}>1$.

\subsection{$a_\mu$ and $\tau\to\mu\gamma$}

Considering only the couplings $\lambda^L$ the effect in $\tau\to\mu\gamma$ is negligibly small. Things get much more interesting if we aim at a simultaneous explanation of the anomalous magnetic moment of the muon. In this case chirally enhanced effects also appear in $\tau\to\mu\gamma$. We have
\begin{equation}
{\rm{Br}}\left[ {\tau  \to \mu \gamma } \right] \geq \frac{{{\alpha}m_\tau ^3}}{{16{\Gamma _\tau }}}\frac{{a_\mu ^2}}{{m_\mu ^2}}{\left| {\frac{{\lambda _{33}^L}}{{\lambda _{32}^L}}} \right|^2}\,.\label{taumugamma}
\end{equation}
Here we set $\lambda^R_{33}=0$. 

Note that ${\rm{Br}}\left( {\tau  \to \mu \gamma } \right)$ can only be enhanced by allowing $\lambda^R_{33}$ to be different from zero, resulting in the $\geq$ sign in \eq{taumugamma}. The result is shown in the right plot of Fig.~\ref{BKtaumuAMM}. Note that $a_\mu$ can only be explained for ${{\lambda _{33}^L}}/{{\lambda _{32}^L}}<0.65$ (at the $2\,\sigma$ level). This is opposite to the case of $b\to s\mu^+\mu^-$ which can only be explained for ${{\lambda _{33}^L}}/{{\lambda _{32}^L}}>1$. Therefore, we conclude that our model can explain out of the three anomalies $R(D^{(*)})$, $b\to s\mu^+\mu^-$ and $a_\mu$ only two simultaneously.

\section{Conclusions and outlook}\label{conclusions}

In this article we proposed a scalar leptoquark model which can give sizable effects on $R(D)$ and $R(D^*)$ without suffering from problems with $b\to s\nu\nu$, $q^2$ distributions in $R(D^{(*)})$, from large couplings in the non-perturbative regime or from tensions with direct LHC searches as it is the case for nearly all other models on the market. Our model predicts sizable branching factions for $b\to s\tau^+\tau^-$ processes (of the order of $10^{-3}$) being directly correlated to $R(D^{(*)})$.

Furthermore, the model can naturally explain $b\to s\mu^+\mu^-$ (including $R(K)$) via a $C_9=-C_{10}$ contribution and therefore also predicts $R(K^*)$ to be significantly below the SM value. In case of a simultaneous explanation of $R(D^{(*)})$ with $b\to s\mu^+\mu^-$ we only get effects in $B\to D^{(*)} \mu\nu/B\to D^{(*)} e\nu$ at the per-mill level, but sizable ones in $b\to s\tau\mu$ processes (depending on only one free parameter), making them potentially observable at LHCb or BELLE II in the near future.

The tension in $a_\mu$ can be explained as well by adding a small right-handed coupling of the $SU(2)_L$ singlet LQ to tops and muons. As a consequence sizable rates for $\tau\to\mu\gamma$ are predicted. Here the dependence on the remaining free parameter is opposite to $b\to s\tau\mu$ excluding a simultaneous explanation of all three anomalies, i.e. out of $R(D^{(*)})$, $b\to s\mu^+\mu^-$ and $a_\mu$ our model can explain any two of them.

We stress that in general our approach of combining the $SU(2)_L$ singlet with the $SU(2)_L$ triplet is the only way of explaining $R(D)$ and $R(D^*)$, without violating $b\to s\nu\nu$ or direct LHC bounds, if the SM is extended by scalar LQs only. However, one can get the same phenomenology in $B$ decays using a vector leptoquark $SU(2)_L$ singlet with hypercharge $2/3$ and couplings to left-handed fermions. In this case, no tuning between couplings is necessary and effects in $b\to s\nu\nu$ are automatically avoided. However, the model is non-renormalizable and while our model with scalar leptoquarks only predicts effects of the order of 1\% in $B_s-\overline{B}_s$ mixing, this effect is much larger for the vector leptoquark~\cite{Barbieri:2015yvd}. Furthermore, adding right-handed couplings, the effect of the vector leptoquark in $a_\mu$ is only enhanced by the bottom mass but not by the top one. 

\begin{figure}[t]
\begin{center}
\begin{tabular}{cp{7mm}c}
\includegraphics[width=0.5\textwidth]{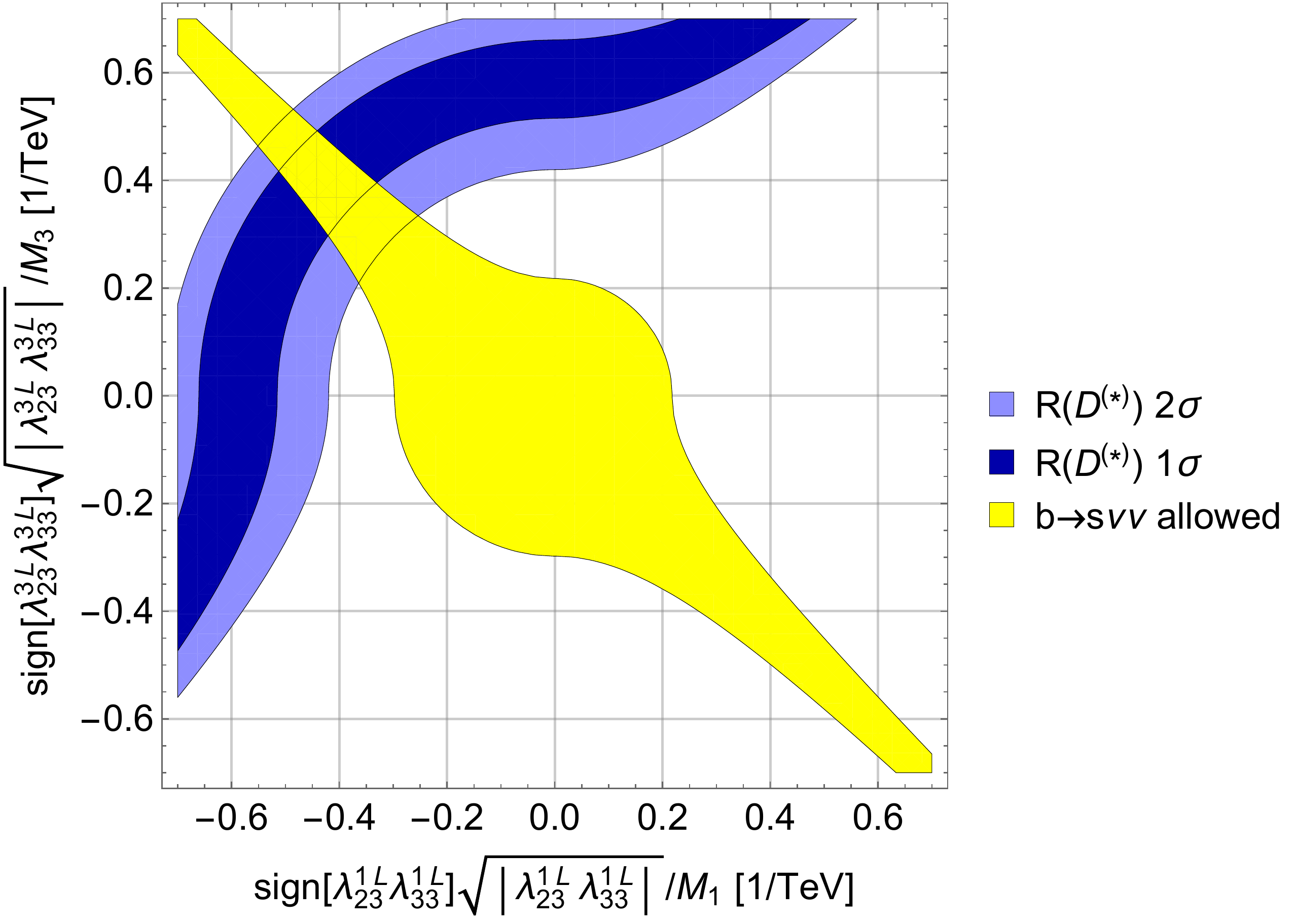}
\end{tabular}
\end{center}
\caption{Allowed regions for $R(D^{(*)})$ and $b\to s\nu\nu$ assuming independent couplings and masses for the leptquark singlet and triplet. Here $M_{1(3)}$ is the mass of $\Phi_1$ ($\Phi_3$).}         
\label{kappa1kappa3}
\end{figure}

In our model we assumed a discrete symmetry between the couplings of the two leptoquarks in order to cancel exactly the effect in $b\to s\nu\nu$. However, even if one disregards this assumption and allows for independent masses and couplings of $\Phi_1$ and $\Phi_3$, a cancellation in $b\to s\nu\nu$ is still possible. In fact, as shown in Fig. \ref{kappa1kappa3} the constraints from $b\to s\nu\nu$ transitions still allow for an explanation of $R(D^{(*)})$ without severe fine-tuning. Therefore, our imposed symmetry does not to be exact in order to provide a valid explanation of the anomalies.

If one allows in addition for direct couplings (not only CKM induced couplings to up-quarks) to first generation quarks, a sizable effect in $K\to\pi\mu\mu/K\to\pi e e$ is possible which could be tested at NA62 or KOTO~\cite{Crivellin:2016vjc} and interesting correlations with $b\to d$ transitions occur. Therefore, a very detailed study of our model is important and promising in order to explore the many interesting effects which can be observed by ongoing and future experiments.

{\it Acknowledgments} --- {\small A.C. thanks Admir Greljo for useful discussions about LHC searches and Michael Spira for useful comments on the manuscript. The work of A.C. and D.M. is supported by an Ambizione Grant of the Swiss National Science Foundation (PZ00P2\_154834).}

{\it Note added} --- After submission of this article, LHCb presented results for the ratio $R(K^*)$~\cite{LHCbseminar2017} which are in agreement with the predictions of our model. Including the new measurement in the global fit to $b\to s\mu^+\mu^-$ data, the significance for NP in $C_9=-C_{10}$ increased to $5.2\,\sigma$~\cite{Capdevila:2017bsm}.

\bibliography{BIB}

\end{document}